\documentclass{optica-article}

\journal{opticajournal} % for journals or Optica Open

\articletype{Research Article}

\usepackage{lineno}
\usepackage{upgreek}
\usepackage{color,soul}
\usepackage{amsmath}
\usepackage{pdfpages}
\usepackage{graphicx}
\usepackage{soul}
\usepackage{parskip}
\usepackage{caption}
\begin{document}

\title{Telecom-to-Visible Quantum Frequency Converter on a Silicon Nitride Chip}

\author{Sidarth Raghunathan,\authormark{1,*} Richard Oliver,\authormark{1} Yun Zhao, \authormark{1} Karl McNulty, \authormark{2} Chaitali Joshi,\authormark{1, \textdagger} Michal Lipson, \authormark{1,2} and Alexander L. Gaeta \authormark{1,2}}

\address{\authormark{1}Department of Applied Physics and Applied Mathematics, Columbia University, New York, NY, 10027 USA \\
\authormark{2}Department of Electrical Engineering, Columbia University, New York, NY, 10027 USA \\
\authormark{\textdagger}Present affiliation: Google Quantum AI, Santa Barbara, 93111, US}

\email{\authormark{*}sar2250@columbia.edu} %% email address is required; see note below about the corresponding author designation

% use {asbstract*} to suppress the copyright line. Copyright information will be added in production

\begin{abstract*} 
Quantum frequency conversion serves a key role in the realization of hybrid quantum networks by interfacing between wavelength-incompatible platforms. Here we present the first quantum frequency converter connecting visible and telecom domains on a silicon nitride (SiN) chip, using Bragg-scattering four-wave mixing to upconvert heralded single photons from 1260 to 698 nm, which covers a 192 THz span. We examine the noise sources in SiN and devise approaches to suppress noise photons at the source and target frequencies to enable measurements at the single-photon level. We demonstrate an on-chip conversion efficiency of 5\% in photon flux and describe design modifications that can be implemented to significantly improve it. Our results pave the way for the implementation of CMOS-compatible devices in quantum networks.

\end{abstract*}

%%%%%%%%%%%%%%%%%%%%%%%%%%  body  %%%%%%%%%%%%%%%%%%%%%%%%%%
\section{Introduction}
In a mature quantum internet, channels that link processor nodes will be required to transport quantum information from site to site with high fidelity \cite{Kimble}. Most candidates for memories and processor nodes in quantum networks have emission wavelengths in the visible and near-infrared portions of the electromagnetic spectrum. However, commercial single-mode optical fiber has propagation loss of $>5$ dB/km at wavelengths below 800 nm, rendering these wavelengths unsuitable for long-distance fiber communications. A crucial feature in the development of long-distance quantum networks will be a method of downconverting photons to telecom wavelengths, where fiber losses of $<0.2$ dB/km can be exploited, while simultaneously preserving their quantum statistics. 
\par

Current state-of-the-art integrated devices for telecom-visible quantum frequency conversion (QFC) use sum- and difference-frequency generation (SFG/DFG) in $\chi^{(2)}$ waveguides, based on periodically-poled lithium niobate (PPLN) \cite{Tanzili, Ikuta, Fejer, Jianwei Pan, Dreau, Schafer, Bersin, Zaske, Maring, Kartik-PPLN}. PPLN-based devices have been shown to convert light between visible and telecom freqencies at the single-photon level from attenuated coherent states \cite{Fejer}, spontaneous parametric down-conversion (SPDC) \cite{Tanzili, Ikuta, Jianwei Pan}, nitrogen and silicon-vacancy centers in diamond \cite{Dreau, Schafer, Bersin}, rare-earth-doped crystals \cite{Maring}, and semiconductor quantum dots \cite{Kartik-PPLN, Zaske}. A promising alternative to QFC based on $\chi^{(2)}$ processes is $\chi^{(3)}$-based Bragg-scattering four-wave mixing (BS-FWM), in which conversion between a signal and idler photon is mediated by two pumps \cite{McKinstrie, Inoue, Raymer}. Since the $\chi^{(3)}$ nonlinearity is present in all materials, a quantum frequency converter based on BS-FWM can be fabricated directly on a monolithic CMOS-compatible platform. In addition, the use of two pump fields provides an additional tunable degree of freedom that enables BS-FWM to be used to readily translate photons separated by MHz to several hundreds of THz; SFG and DFG by contrast can typically only accommodate large spectral separations. Furthermore, BS-FWM allows flexibility in the choice of pump wavelengths since BS-FWM conversion depends only on the pump frequency separation. This existence of multiple pump wavelength configurations for a single signal/idler pair allows one to choose from commercially mature laser wavelengths and to position the pumps in a way that optimizes the noise performance of the device. \par

BS-FWM-based QFC has been shown with efficiencies exceeding 95\% in optical fibers with signal and idler wavelengths separated by a few nanometers in the telecom band \cite{Stephane, Eggleton}. It has also been achieved using atomic transitions in rubidium vapor \cite{Radnaev, Figueroa} to connect a rubidium quantum memory with telecom wavelengths for optical fiber transmission. In integrated platforms, BS-FWM has been shown to translate emission from a quantum dot across 1 nm with 12\% efficiency \cite{Srinivasan QD}. In spite of these demonstrations, wide-spanning on-chip BS-FWM frequency conversion has yet to be shown in the single-photon regime. The largest signal-idler separation demonstrated with classical BS-FWM was 115 THz, between 980 and 1550 nm \cite{Srinivasan}. However, excessive contamination by noise photons from the nearby pumps made this device challenging for QFC.\par

In this work, we show conversion across 192 THz with single photons in a microresonator and perform a characterization of noise sources in silicon nitride. Silicon nitride has been a challenging platform for quantum photonics experiments spanning visible and telecom bands since it can exhibit strong fluorescence at visible wavelengths, and SFWM noise prevents the placement of the signal or target fields near either of the pumps, where phase-matching for BS-FWM is most convenient. We reduce these sources of noise to acceptable levels by placing the pumps in the normal group-velocity dispersion regime, utilizing a large 40-THz frequency separation between each of the pumps and its nearest signal/idler field, and ensuring the near-visible pump is red-detuned from the idler resonance. We further optimize the noise properties by adjusting the relative powers of the pumps, demonstrating that silicon nitride can serve as a low-noise platform for QFC across widely separated wavelength bands.
 This is, to the best of our knowledge, the widest-spanning demonstration of four-wave-mixing-based QFC in any solid-state platform, and the first one linking telecom and visible domains.\par

\section{Theory of Cavity-Confined BS-FWM}
The conversion efficiency $\eta$ of BS-FWM in a resonator can be derived using a four-mode model (see Appendix A for details). After optimizing the resonance detunings of both pumps fields, the efficiency (plotted in Fig. \ref{fig:noise_schematic}a) is given by,

\begin{equation}
  \eta =
    \begin{cases}
       \displaystyle{\frac{\theta_\mathrm{s} \theta_\mathrm{i}}{\alpha_\mathrm{s} \alpha_\mathrm{i}} \frac{4\mathcal{C}}{(1+\mathcal{C})^2}} & \text{for $ \mathcal{C} <$ 1,}\\ \\
      \displaystyle{\frac{\theta_\mathrm{s} \theta_\mathrm{i}}{\alpha_\mathrm{s}\alpha_\mathrm{i}}}  & \text{for $ \mathcal{C} >$ 1,}\\
    \end{cases}
\end{equation} 

\noindent where 
\begin{equation}
\mathcal{C}=\frac{16\gamma_\mathrm{s} \gamma_\mathrm{i} L^2 P_{1}P_{2}}{\alpha_\mathrm{s} \alpha_\mathrm{i}},
\end{equation}

\noindent and $\theta_{s, i}$ is the coupling coefficient at the signal (idler) wavelength, $\alpha_{s, i}$ is the roundtrip loss coefficient at the signal (idler) wavelength, $P_{1,2}$ refers to the intracavity power of each pump, $L$ is the resonator circumference, and $\gamma_{s, i}$ is the nonlinear parameter at signal (idler) wavelength. The cooperativity $\mathcal{C}$ is given by the ratio of the nonlinear coupling rate to the linear dissipative rate. The conversion efficiency exhibits two regimes governed by the magnitude of the cooperativity. When $\mathcal{C}>1$, known as the overpumped regime, the maximum efficiency saturates at an upper bound given by the ratio of the resonator-bus coupling coefficients at the signal and idler wavelengths to their roundtrip loss coefficients \cite{Sipe, Yun Theory}. This is the regime in which we choose to operate. \par

\section{Device Design}
\begin{figure}[ht!]
\centering\includegraphics[width=12cm]{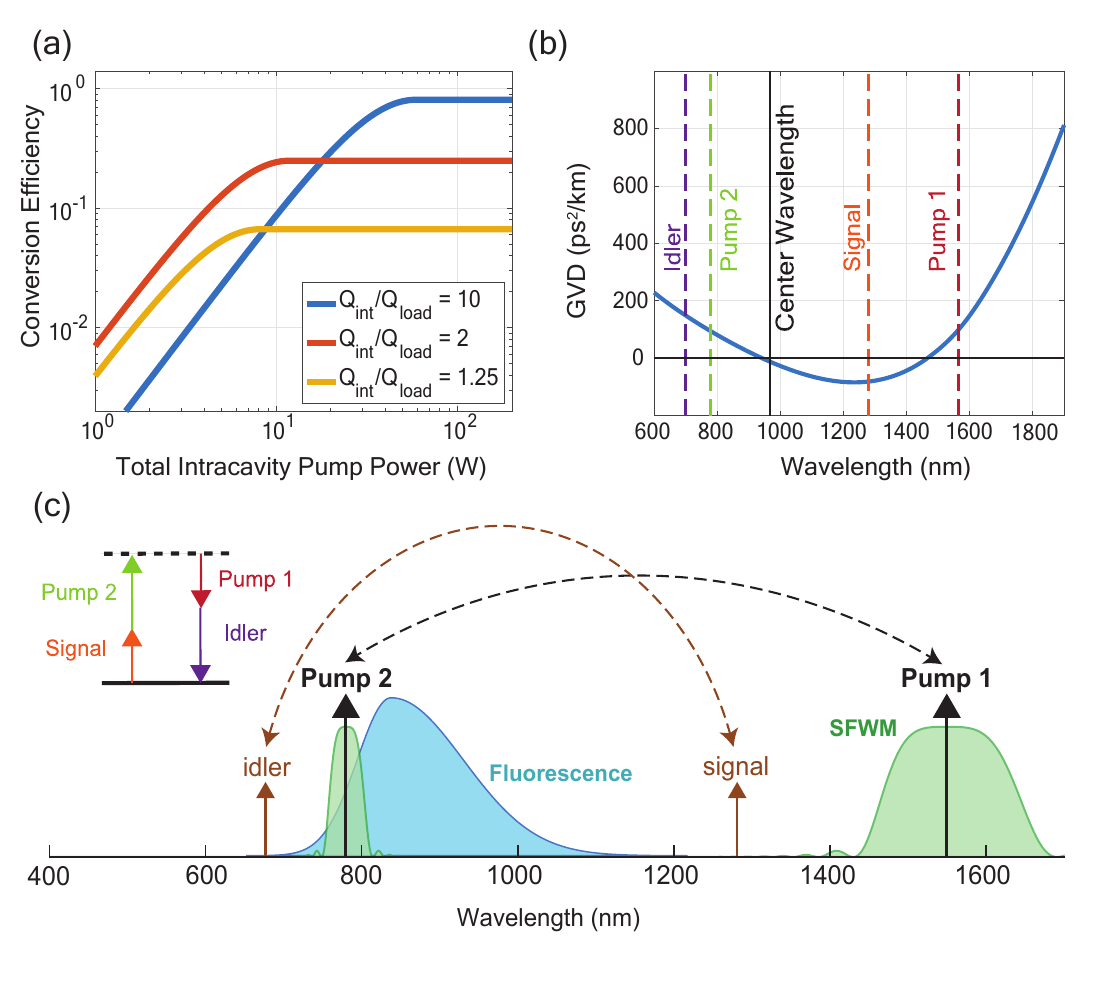}
\caption{(a) BS-FWM conversion efficiency as a function of pump power in a resonator based on coupled-mode theory assuming overcoupled, critically coupled, and undercoupled signal/idler. (b) Simulated group-velocity dispersion (GVD) of our QFC device. (c) Depiction of the various noise sources in silicon nitride. Inset on left shows the energy diagram for the BS-FWM process.}
\label{fig:noise_schematic}
\end{figure}

The desired BS-FWM interaction must be phase-matched to optimize the efficiency of the QFC process. Here we set the target idler wavelength to 698 nm, corresponding to the strontium clock transition, which could serve as a quantum memory due to its millihertz linewidth. We choose a signal wavelength that lies in the telecom O-band (1260-1360 nm), where loss coefficients of $<0.5$ dB/km are standard for single-mode fibers. Phase-matching is satisfied by situating the zero group-velocity dispersion (GVD) point near the center wavelength of the BS-FWM process. This can be achieved by designing the cross-section of the waveguide to be 730 $\times$ 1300 nm and using the TM mode, which satisfies this constraint and also maintains both pumps in a region of normal GVD where spontaneous pair generation is not phase-matched (see Fig. \ref{fig:noise_schematic}b). \par

Further optimization of the QFC process is achieved by operating in the higher overcoupled regime at both the signal and idler frequencies to maximize the conversion efficiency (see Eq. 1). This is challenging at visible wavelengths since at these higher frequencies the mode becomes tightly confined, which reduces the evanescent overlap between the mode in the bus waveguide and the mode in the microring. We address this issue by utilizing a pulley-coupler design in which the bus waveguide wraps around the ring to increase the interaction length \cite{Adibi, Srinivasan}. We then choose a bus waveguide width that ensures that the field in the bus waveguide and the field in the ring remain in-phase as they propagate at the idler wavelength, so that the coupling coefficient builds up constructively along the length of the pulley coupler. In our experiment, we find that the large interaction length required to highly overcouple the short-wavelength resonances resulted in a reduced intrinsic quality factor for the telecom-band resonances, due to increased interaction of the hybrid ring-bus mode with the waveguide sidewalls in the coupling region. As a result, we opted for a device that was less strongly overcoupled at shorter wavelengths, in order to preserve the intrinsic quality factors at telecom wavelengths. For the device used in this experiment, we measure loaded quality factors of $1.5 \times 10^6$ at 780 nm, $4.5 \times 10^5$ at 1283 nm, and $2.8 \times 10^5$ at 1548 nm, as shown in Fig. \ref{fig:Figure 2}a. \par

\section{Results and Noise Characterization}
\begin{figure}[ht!]
\centering\includegraphics[width=12cm]{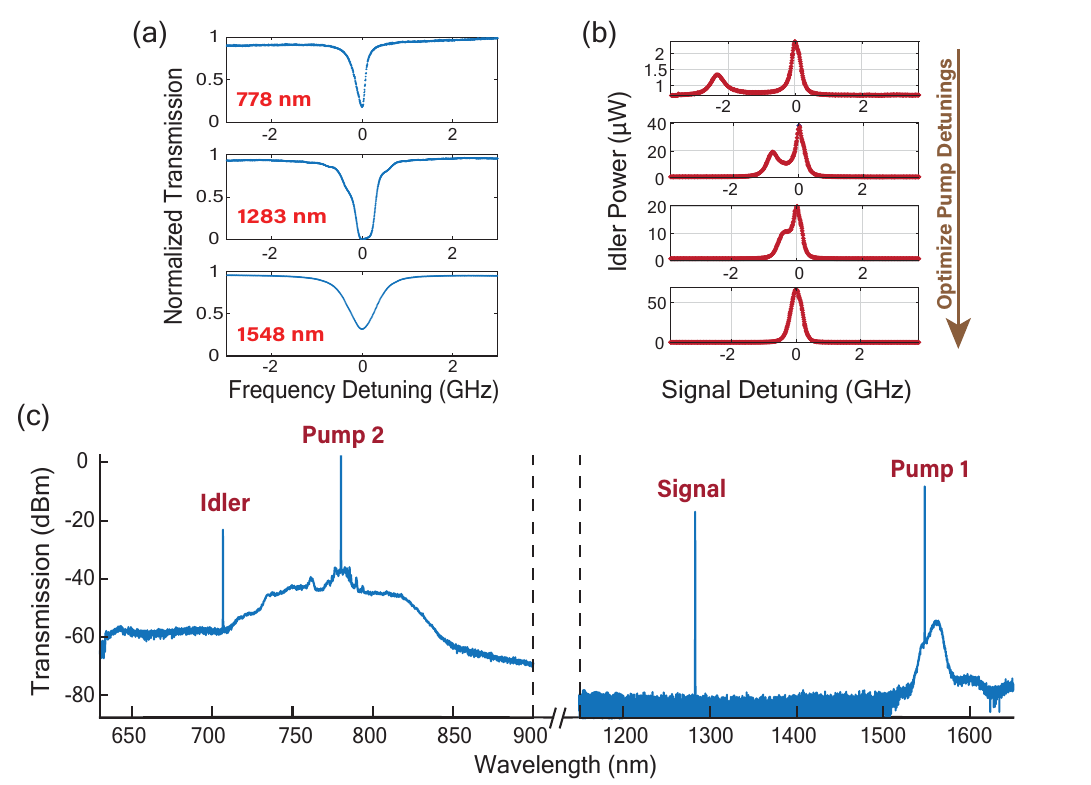}
\caption{(a) Transmission scans of cavity resonances at the signal and pump wavelengths. (b) Rabi-like splitting is induced at the idler resonance during frequency conversion due to strong nonlinear coupling between the signal and idler. Different traces correspond to different pump detunings. (c) Trace obtained from the optical spectrum analyzer with strong coherent state input at the signal wavelength.}
\label{fig:Figure 2}
\end{figure}

We investigate conversion in the classical regime by injecting an O-band coherent state into the microresonator. The pumps are then iteratively tuned into their respective resonances, due to cavity hysteresis arising from Kerr and thermal nonlinearities. A maximum conversion efficiency of 6\% is observed when the signal is at 1283 nm and the idler is at 704 nm (Fig. \ref{fig:Figure 2}c).  By tuning the signal to 1260 nm, conversion to 698 nm is achieved with 5\% efficiency. Typically, chip-integrated BS-FWM demonstrations are plagued by parasitic nonlinear processes which require specialized suppression mechanisms to avoid major reductions in conversion efficiency \cite{Bell, Heuck, Lacava}. In our experiment, we keep the wavelength separation between each field sufficiently large such that the phase-matching conditions for the desired BS-FWM process and the parasitic processes are significantly different. As a result, we do not observe generation of light at any frequency other than that of the idler, validating the applicability of our four-mode model. We also observe Rabi-like splitting in the idler mode induced by strong nonlinear coupling (see Fig. \ref{fig:Figure 2}b) \cite{Hong Tang, Sven}, which is a signature of the overpumped regime in which the conversion efficiency upper bound is achievable. By scanning the signal across the cavity resonance, we infer that it is critically coupled (Fig. \ref{fig:Figure 2}a). While we cannot determine the coupling at the idler wavelength since a tunable laser at the idler frequency was not available, it is likely that the idler resonance is undercoupled, which according to Eq. 1 would limit the theoretically achievable conversion efficiency in this device to less than 25\%. \par

A key aspect for high-fidelity QFC is ensuring that there are no sources of noise photons that contribute at the signal and idler frequencies. We identify two primary sources of noise photons in silicon nitride to be fluorescence and spontaneous four-wave mixing (SFWM), both of which are illustrated in Fig. \ref{fig:noise_schematic}c. Fluorescence is primarily an issue when the waveguide is pumped at wavelengths below 1.1 $\mathrm{\upmu}$m \cite{Yun Visible}. The origin of fluorescence noise in silicon nitride is not fully understood. Two possibilities are that it arises from silicon nanoparticle defects pumped above their bandgap energy \cite{silicon_nanoparticles} or that it is due to bandtail transitions intrinsic to silicon nitride \cite{band_tail}. In our experiment, we find that fluorescence is concentrated on the red side of the 780-nm pump up to about 1100 nm, which corresponds to the bandgap of silicon. By placing both pumps on the long wavelength side of the idler, we can mitigate contamination from fluorescence noise. The second major source of noise photons from spontaneous four-wave mixing (SFWM) is localized around the pump wavelength and is more pronounced and broader bandwidth in the anomalous-GVD regime. By placing both pumps in the normal-GVD regime and utilizing a large, 40-THz separation between the signal or idler and the nearest pump, we suppress contamination from SFWM. SFWM scales quadratically with pump power and is co-polarized with the pump, whereas fluorescence should at low powers scale linearly with pump power and be unpolarized \cite{Ryota, Yun Visible}. Spontaneous Raman scattering, if present, should also scale linearly with pump power but be co-polarized with the pump \cite{McKinstrie}. This offers an experimental means of distinguishing these different noise sources at the source and target wavelengths in order to characterize the noise properties of the device. \par

\begin{figure}[ht!]
\centering\includegraphics[width=10cm]{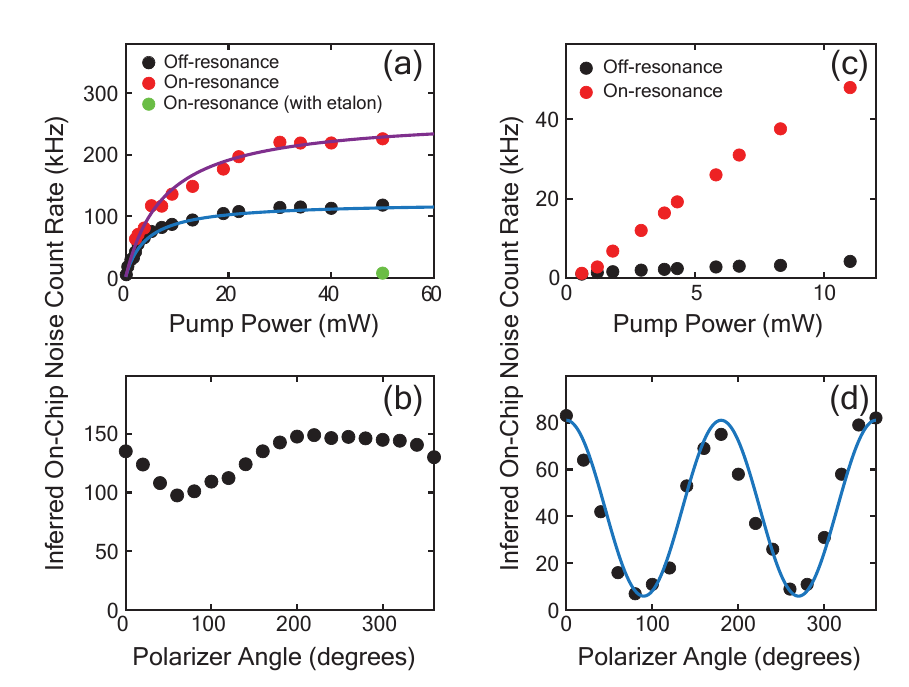}
\caption{Noise counts at (left) 700 nm, and (right) 1260 nm as functions of pump power (a,c) and polarization (b,d). Losses after the chip, including the quantum efficiencies of the detectors, are corrected for.}
\label{fig:noise_characterization}
\end{figure}

In the absence of an injected signal, we characterize the pump-induced noise at the idler wavelength with both on- and off-resonance pumps using a single-photon avalanche detector (SPAD). We place a 10-nm bandpass filter centered at 700 nm (Thorlabs FBH700-10) immediately after the chip and measure resonant noise counts of 55 kHz generated by the 780-nm pump (which we infer to be  220 kHz on-chip after correcting for linear losses after the chip and the quantum efficiency of the SPAD). These counts show a weak pump-polarization dependence and have a pump power scaling that is linear at low powers and saturates at high powers [Fig. \ref{fig:noise_characterization}a-b], which is characteristic of fluorescence. Since this noise is broadband, we can effectively suppress it by replacing the bandpass filter with an etalon of finesse 40, which lowers the idler transmission by 30\% but reduces the noise counts by 15 dB. \par

In order to characterize the noise at the signal frequency, we use a series of fiber Bragg gratings to isolate a 1-nm window at 1260 nm and observe noise counts in this range using a superconducting nanowire single-photon detector (SNSPD). We find that the 1550-nm pump generates a negligible number of noise photons, but counterintuitively we do see substantial noise generated by the 780-nm pump. This noise is co-polarized with the pump and scales linearly with pump power (Fig. \ref{fig:noise_characterization}c-d). Noise photons of this nature could arise from high-frequency-phonon-induced spontaneous Raman scattering and material-surface-induced $\chi^{(2)}$ nonlinear processes. Since the telecom band noise photons are primarily resonant with the cavity, a different noise suppression approach must be implemented. To this end, we utilize the extra degree of freedom provided by the dual-pump nature of BS-FWM to suppress the noise. Since the conversion efficiency depends only on the product of the pump powers (Eq. 1), we operate with a pump power imbalance by reducing the power of the 780-nm pump and compensating with a higher power at 1550 nm. We find that by utilizing an on-chip pump power of 90 mW at 1550 nm but just 4 mW at 780 nm, we measure a total noise count rate of $<3$ kHz at the idler wavelength when both pumps are resonant, while still retaining a large enough pump-power product to keep us in the overpumped regime so that we do not sacrifice conversion efficiency. \par

\begin{figure}[ht!]
\centering\includegraphics[width=12cm]{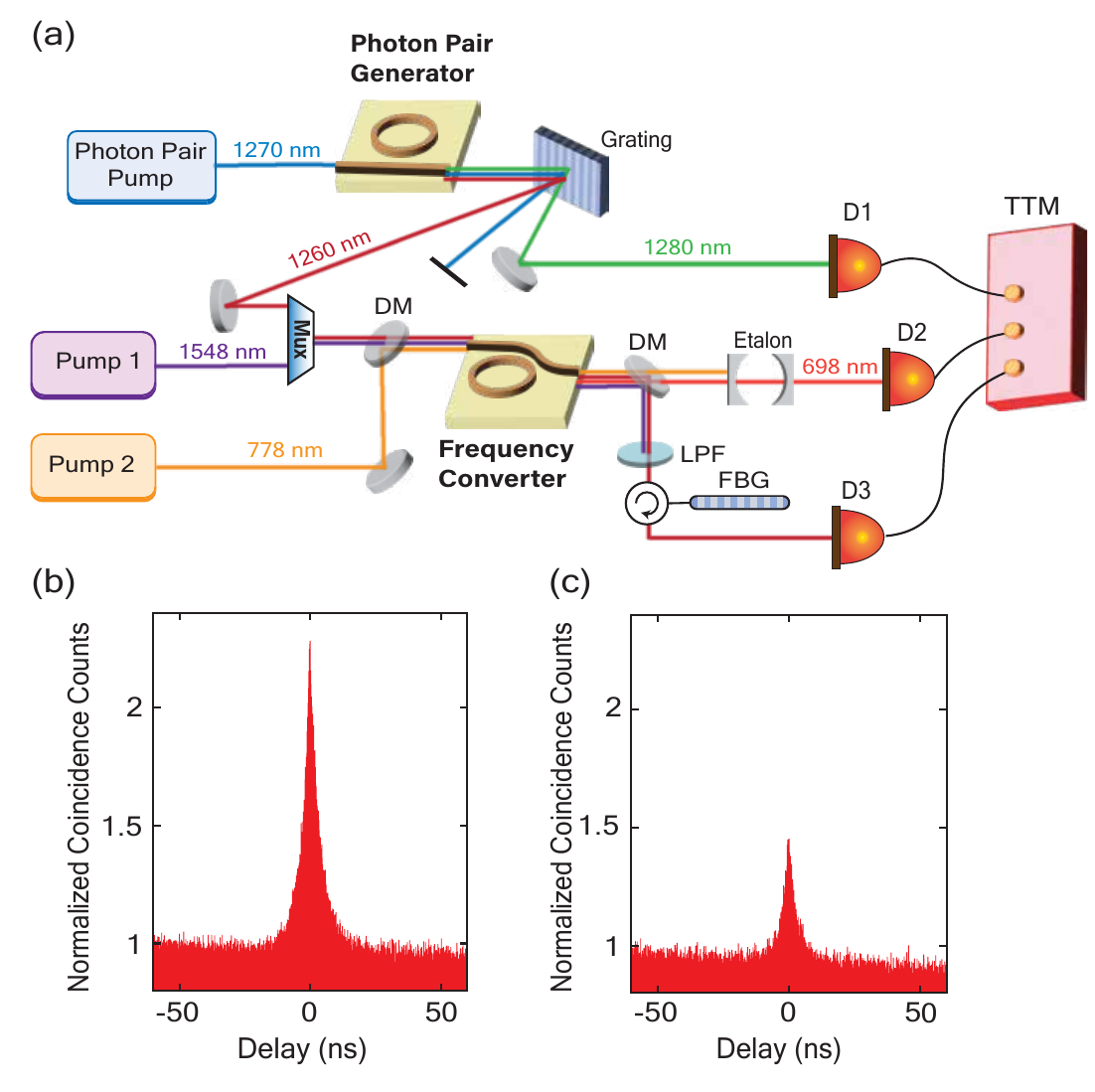}
\caption{(a) Schematic of experimental setup. Photon pairs are generated in a separate SiN microresonator. One member of the pair is routed directly to an SNSPD (D1), while the other is sent to the QFC chip where it is upconverted to 698 nm and detected by a SPAD (D2). The unconverted signal is sent to another SNSPD (D3). DM=dichroic mirror, LPF=low-pass filter, FBG=fiber Bragg grating, TTM=time-tagging module. (b-c) Delay histogram showing coincidence counts between (b) detectors D1 and D3 and (c) detectors D1 and D2.}
\label{fig:experimental_setup}
\end{figure}

Finally, we investigate the QFC process with single-photons injected at the signal wavelength. As shown in Fig. \ref{fig:experimental_setup}a, the signal photons are generated by pumping another SiN micoresonator at 1270 nm to generate photon pairs via SFWM at 1260 and 1280 nm. The generated photon pairs are separated using a diffraction grating and narrowly filtered by propagating them over 6 meters in free space, which provides $>100$ dB rejection at the pump wavelength. The 1260-nm photon is then routed to the QFC chip, where it is upconverted to 698 nm. In order to eliminate noise photons arising from spontaneous Raman scattering in the input fibers, both BS-FWM pumps are coupled out of fiber and filtered in free space using a narrowband filter, before they are combined using a dichroic mirror and sent onto the QFC chip. We send each photon to a separate detector and obtain a delay histogram of coincidence counts between the heralding and frequency-converted heralded photons (Fig. \ref{fig:experimental_setup}b-c). A clear coincidence peak is observed both before and after conversion, indicating the viability of this frequency conversion device in a quantum network. The reduction in visibility is due to the limited conversion efficiency, which can be improved by more strongly overcoupled signal and idler resonances.

\section{Conclusion}

 We have developed a CMOS-compatible system for photon-pair generation and QFC that can translate quantum photonic states between the red portion of the visible spectrum with the near-infrared and telecom bands. In addition to the telecom-visible conversion shown here, such a scheme can also be used to interface between rubidium and strontium atomic species by swapping the 780-nm pump with the signal. Our system overcomes the broadband noise generated by the 780-nm pump by operating at a power imbalance between the 780 and 1550-nm pumps, enabling us to recover low-noise operation without sacrificing conversion efficiency. We further note that fluorescence-induced noise is not a fundamental constraint and can be eliminated by using nitrogen-rich SiN \cite{nitrogen_rich} or in stoichiometric SiN by performing rapid thermal annealing in argon gas \cite{thermal_annealing}. We expect that further optimization of the ring-bus coupling region will allow for conversion efficiencies that approach unity. \par

\section{Funding}
This research was supported in part by the National Science Foundation PHY-2110615 and OSI-2410725, and the Department of Energy, Office of Science, National Quantum Information Science Research Centers, Co-design Center for Quantum Advantage (C2QA).

\section{Acknowledgment}
This work was performed in part at the Cornell NanoScale Facility, a member of the National Nanotechnology Coordinated Infrastructure (NNCI), which is supported by the NSF. Author acknowledges useful discussions with Ryota Katsumi.

\section{Disclosures}
The authors declare no conflicts of interest. \par

Please see Supplement 1 for supporting content.

%%%%%%%%%%%%%%%%%%%%%%% References %%%%%%%%%%%%%%%%%%%%%%%%%

 \pagebreak

\huge \begin{center} Supplementary Material \end{center} 

\setcounter{section}{0} 
\normalsize \section{Derivation of Conversion Efficiency for Cavity-Confined BS-FWM}

We can write the coupled amplitude equations for conversion between the intracavity signal field $C$ and idler field $D$ mediated by two pump fields $A$ and $B$ in matrix form \cite{Yun Thesis}:

\begin{equation*}
    t_R \frac{d}{dt} \begin{bmatrix}
        C\\ D
    \end{bmatrix} = \begin{bmatrix}
        -\frac{\alpha_{C}}{2} - i \Delta_{C} + 2i \gamma_C L (|A|^2 + |B|^2) & 2i\gamma_C L A B^{*} \\
        2i \gamma_D L A^{*} B & -\frac{\alpha_{D}}{2} - i\Delta_{D} + 2i \gamma_D L(|A|^2+|B|^2) 
    \end{bmatrix} \begin{bmatrix}
        C \\ D 
    \end{bmatrix} + \begin{bmatrix}
        \sqrt{\theta_{C}} & 0 \\ 0 & \sqrt{\theta_{D}} 
    \end{bmatrix} \begin{bmatrix}
        C_{in} \\ D_{in} 
    \end{bmatrix} 
\end{equation*}

\begin{equation}
    \equiv M\begin{bmatrix}
        C \\ D
    \end{bmatrix} + K \begin{bmatrix}
    C_{in} \\ D_{in}
    \end{bmatrix}
    \label{eq:S1}
\end{equation}

\noindent where $\gamma_{C (D)} $ is the nonlinear parameter at the signal (idler) wavelength, $L$ is the circumference of the resonator, $t_R$ is the roundtrip time of the resonator, $\alpha_{C (D)}$ is the total loss coefficient (i.e. corresponding to the loaded resonance linewidth) at the signal (idler) wavelengths, $\Delta_{C (D)}$ is the resonance detuning of the signal (idler) field, and $\theta_{C (D)}$ is the coupling coefficient at the signal (idler) wavelength. \par

\noindent In order to extract the signal and idler fields coupled out of the ring into the bus waveguide, we can use the following beamsplitter relation:

\begin{equation}
    \begin{bmatrix}
        C_{out} \\ D_{out}
    \end{bmatrix} = K \begin{bmatrix}
        C \\ D
    \end{bmatrix} - \sqrt{1-K^2}\begin{bmatrix}
        C_{in} \\ D_{in}
    \end{bmatrix}
    \label{eq:S2}
\end{equation}

\noindent In steady state, we can set the left-hand side of Eq. \ref{eq:S1} to zero:
\begin{equation}
    \begin{bmatrix}
        C \\ D
    \end{bmatrix} = -M^{-1} K \begin{bmatrix}
        C_{in} \\ D_{in}
    \end{bmatrix}
\end{equation}

\noindent Define $\zeta_{C (D)} \equiv -\alpha_{C (D)}/2 + 2i\gamma_{C(D)} L (|A|^2 + |B|^2)$. Then,
\begin{equation}
    \mathrm{det(M)} = (\zeta_{C} - i\Delta_{C})(\zeta_{D} - i\Delta_{D}) + 4\gamma_C \gamma_D L^2 |A|^2 |B|^2
\end{equation} 
\begin{equation}
    \begin{bmatrix}
        C \\ D
    \end{bmatrix} = 
    -\frac{1}{\mathrm{det(M)}} \begin{bmatrix}
        \sqrt{\theta_{C}} (\zeta_D - i\Delta_D) & -\sqrt{\theta_D} 2i\gamma_C L AB^{*} \\
        -\sqrt{\theta_{C}} 2i \gamma_D L A^{*}B & \sqrt{\theta_D}(\zeta_C - i\Delta_C)
    \end{bmatrix} \begin{bmatrix}
        C_{in} \\ D_{in} 
    \end{bmatrix}
    \label{eq:S5}
\end{equation} 

\noindent Then, combining Eq. \ref{eq:S5} with the beamsplitter relation in Eq. \ref{eq:S2}, we can obtain a direct relationship between the input and output fields in the bus waveguide:

\begin{equation} 
    \begin{bmatrix}
        C_{out} \\ D_{out}
    \end{bmatrix} = 
    -\Bigg( \frac{1}{\mathrm{det(M)}} \begin{bmatrix}
        \theta_{C} (\zeta_D - i\Delta_D) & -\sqrt{\theta_C \theta_D} 2i\gamma_C L AB^{*} \\
        \sqrt{\theta_{C}\theta_D} 2i \gamma_D L A^{*}B & \theta_D(\zeta_C - i\Delta_C)
    \end{bmatrix} + \begin{bmatrix}
        \sqrt{1-\theta_C} & 0 \\
        0 & \sqrt{1-\theta_D} \\
    \end{bmatrix} \Bigg) \begin{bmatrix} 
        C_{in} \\ D_{in} 
    \end{bmatrix}
    \label{S5}
\end{equation}

\noindent When there is no input idler field (i.e. $D_{in} = 0$), we can write the efficiency as: 

\begin{equation}
    \eta \equiv \bigg|\frac{D_{out}}{C_{in}}\bigg|^2 = \frac{4\gamma_D^2 L^2 \theta_C \theta_D |A|^2 |B|^2}{\Big|(\zeta_C - i\Delta_C)(\zeta_D - i\Delta_D) + 4\gamma_C \gamma_D L^2|A|^2 |B|^2 \Big|^2}
    \label{S6}
\end{equation} 

\noindent There is an exceptional point equal to the ratio of the nonlinear coupling rate to the dissipative rate of the resonator which we define as the cooperativity parameter $\mathcal{C}$. When it equals unity, in the limits where signal and idler are close together in frequency, we transition from a regime where the eigenvalues of the transfer matrix relating the input and output fields in Eq. \ref{S5} have two degenerate eigenvalues to a regime where both of their eigenvalues are nondegenerate. Its value is given by: \par
\begin{equation*}
    \mathcal{C} = \sqrt{\frac{16 \gamma_C \gamma_D L^2}{\alpha_C \alpha_D}}
\end{equation*}

\noindent We call the regime where $\mathcal{C}<1$ the underpumped regime, and the regime where $\mathcal{C}>1$ as the overpumped regime \cite{Yun Theory}. The two nondegenerate eigenvalues in the overpumped regime correspond to a Rabi-like splitting induced by strong nonlinear coupling between the signal and idler fields. 

\noindent It can be verified from equation \ref{S6} that for fixed pump power values, $\eta$ is maximized for detuning values of:

\begin{equation}
    \Delta_C = 2\gamma_C L (|A|^2 + |B|^2) \hspace{2cm} \Delta_D = 2\gamma_D L (|A|^2+|B|^2)
    \label{S7}
\end{equation}
in the underpumped regime, and 
\begin{equation}
    \Delta_D = 2 \gamma_D L(|A|^2+|B|^2) - \frac{\alpha_D}{2} \sqrt{r-1}
    \label{S8}
\end{equation}
\begin{equation}
    \Delta_C = \frac{\alpha_C}{\alpha_D} (\Delta_D - 2 \gamma_D L (|A|^2+|B|^2)) + 2 \gamma_C L (|A|^2+|B|^2)
    \label{S9}
\end{equation}
in the overpumped regime. \par

\noindent Inserting equations \ref{S7}-\ref{S9} into equation \ref{S6} gives us a maximum conversion efficiency of:
\begin{equation}
    \eta = \frac{\theta_C \theta_D}{\alpha_C \alpha_D} \frac{4r}{(1+r)^2} 
\end{equation}
in the underpumped regime, and
\begin{equation}
    \eta = \frac{\theta_C \theta_D}{\alpha_C \alpha_D}
\end{equation}
in the overpumped regime.


\begin{thebibliography}{1}
\newcommand{\enquote}[1]{``#1''}

\bibitem{Kimble}
H. Kimble, "The quantum internet," Nature \textbf{453}, 1023–1030 (2008).

\bibitem{Tanzili}
S. Tanzilli, W. Tittel, M. Halder, O. Alibart, P. Baldi, N. Gisin, and H. Zbinden, "A Photonic Quantum Information Interface," Nature \textbf{437}, 116–120 (2005).

\bibitem{Ikuta}
R. Ikuta, Y. Kusaka, T. Kitano, H. Kato, T. Yamamoto, M. Koashi, N. Imoto, "Wide-band quantum interface for visible-to-telecommunication wavelength conversion." Nat Commun \textbf{2}, 537 (2011). 

\bibitem{Jianwei Pan}
X. Wang, X. Jiao, B. Wang, Y. Liu, X.-P. Xie, M.-Y. Zheng, Q. Zhang and J.-W. Pan, "Quantum frequency conversion and single-photon detection with lithium niobate nanophotonic chips." npj Quantum Inf 9, 38 (2023).

\bibitem{Fejer}
J. S. Pelc, L. Ma, C. R. Phillips, Q. Zhang, C. Langrock, O. Slattery, X. Tang, and M. M. Fejer, "Long-wavelength-pumped upconversion single-photon detector at 1550 nm: performance and noise analysis," Opt. Express \textbf{19}, 21445-21456 (2011)

%\bibitem{Pelc}
%J. S. Pelc, L. Yu, K. De Greve, P. L. McMahon, C. M. Natarajan, V. Esfandyarpour, S. Maier, C. Schneider, Martin Kamp, S. Höfling, R. H. Hadfield, A. Forchel, Y. Yamamoto, and M. M. Fejer, "Downconversion quantum interface for a single quantum dot spin and 1550-nm single-photon channel," Opt. Express \textbf{20}, 27510-27519 (2012).

\bibitem{Bersin}
E. Bersin, M. Sutula, Y. Q. Huan, A. Suleymanzade, D. R. Assumpcao, Y.-C. Wei, P.-J. Stas, C. M. Knaut, E. N. Knall, C. Langrock, N. Sinclair, R. Murphy, R. Riedinger, M. Yeh, C. J. Xin, S. Bandyopadhyay, D. D. Sukachev, B. Machielse, D. S. Levonian, M. K. Bhaskar, S. Hamilton, H. Park, M. Loncar, M. M. Fejer, P. B. Dixon, D. R. Englund, M. D. Lukin, "Telecom Networking with a Diamond Quantum Memory," PRX Quantum \textbf{5}, 010303 (2024).

\bibitem{Schafer}
M. Schäfer, B. Kambs, D. Herrmann, T. Bauer, C. Becher,"Two-Stage, Low Noise Quantum Frequency Conversion of Single Photons from Silicon-Vacancy Centers in Diamond to the Telecom C-Band," Adv. Quantum Technol. 2300228 (2023).

\bibitem{Dreau}
A. Dréau, A. Tchebotareva, A. El Mahdaoui, C. Bonato, and R. Hanson, "Quantum Frequency Conversion of Single Photons from a Nitrogen-Vacancy Center in Diamond to Telecommunication Wavelengths," Phys. Rev. Applied \textbf{9}, 064031 (2018).

\bibitem{Maring}
Maring, N., Farrera, P., Kutluer, K. et al. "Photonic quantum state transfer between a cold atomic gas and a crystal," Nature \textbf{551}, 485–488 (2017).

\bibitem{Kartik-PPLN}
M. Rakher, L. Ma, O. Slattery, X. Tang and K. Srinivasan, "Quantum transduction of telecommunications-band single photons from a quantum dot by frequency upconversion," Nature Photon \textbf{4}, 786–791 (2010).

\bibitem{Zaske}
S. Zaske, A. Lenhard, C. A. Keßler, J. Kettler, C. Hepp, C. Arend, R. Albrecht, W.-M. Schulz, M. Jetter, P. Michler, and C. Becher, "Visible-to-Telecom Quantum Frequency Conversion of Light from a Single Quantum Emitter," Phys. Rev. Lett. \textbf{109}, 147404 (2012).

\bibitem{McKinstrie}
C. J. McKinstrie, J. D. Harvey, S. Radic, and M. G. Raymer, "Translation of quantum states by four-wave mixing in fibers." Opt. Express \textbf{13}, 9131 (2005).

\bibitem{Inoue}
K. Inoue, "Tunable and selective wavelength conversion using fiber four-wave mixing with two pump lights," IEEE Photonics Technol. Lett. 6, 1451–1453 (1994)

\bibitem{Raymer}
H. J. McGuinness, M. G. Raymer, C. J. McKinstrie, and S. Radic, "Quantum Frequency Translation of Single-Photon States in a Photonic Crystal Fiber,"
Phys. Rev. Lett. \textbf{105}, 093604 (2010); Erratum Phys. Rev. Lett. 105, 119901 (2010).

\bibitem{Eggleton}
A. Clark, S. Shahnia, M. Collins, C. Xiong, and B. Eggleton, "High-efficiency frequency conversion in the single-photon regime," Opt. Lett. \textbf{38}, 947-949 (2013).

\bibitem{Stephane}
S. Clemmen, A. Farsi, S. Ramelow, and A. L. Gaeta, "Ramsey Interference with Single Photons," 
Phys. Rev. Lett. \textbf{117}, 223601 (2016).

\bibitem{Radnaev}
A. G. Radnaev, Y. O. Dudin, R. Zhao, H. H. Jen, S. D. Jenkins, A. Kuzmich and T. A. B. Kennedy, "A quantum memory with telecom-wavelength conversion," Nature Phys \textbf{6}, 894–899 (2010).

\bibitem{Figueroa}
D. Du, L. Castillo-Veneros, D. Cottrill, G.-D. Cui, G. Bello, M. Flament, P. Stankus, D. Katramatos, J. Martínez-Rincón, E. Figueroa, "A long-distance quantum-capable internet testbed," 	ArXiv:2101.12742 (2021).

\bibitem{Srinivasan QD}
A. Singh, Q. Li, S. Liu, Y. Yu, X. Lu, C. Schneider, S. Höfling, J. Lawall, V. Verma, R. Mirin, S. W. Nam, J. Liu, and K. Srinivasan, "Quantum frequency conversion of a quantum dot single-photon source on a nanophotonic chip," Optica \textbf{6}, 563-569 (2019).

\bibitem{Srinivasan}
Q. Li, M. Davanço, and K. Srinivasan, "Efficient and low-noise single-photon-level frequency conversion interfaces using silicon nanophotonics," Nature Photon \textbf{10}, 406–414 (2016).

\bibitem{Yun Visible}
Y. Zhao, X. Ji, B.Y. Kim, P. Donvalkar, J. Jang, C. Joshi, M. Yu, C. Joshi, R. Domeneguetti, F. Barbosa, P. Nussenzveig, Y. Okawachi, M. Lipson, and A. L. Gaeta, "Visible nonlinear photonics via high-order-mode dispersion engineering," Optica \textbf{7}, 135-141 (2020).

\bibitem{Sipe}
Z. Vernon, M. Liscidini, and J.E. Sipe, "Quantum frequency conversion and strong coupling of photonic modes using
four-wave mixing in integrated microresonators," Phys. Rev. A \textbf{94}, 023810 (2016).

\bibitem{Yun Theory}
Y. Zhao, J. Jang, Y. Okawachi, and A. L. Gaeta, "Theory of $\chi$(2)-microresonator-based frequency conversion," Opt. Lett. \textbf{46}, 5393-5396 (2021)

\bibitem{Bell}
B. Bell, C. Xiong, D. Marpaung, C. J. McKinstrie, and B. J. Eggleton, "Uni-directional wavelength conversion in silicon using four-wave mixing driven by cross-polarized pumps," Opt. Lett. \textbf{42}, 1668 (2017).

\bibitem{Heuck}
M. Heuck, J. G. Koefoed, J. B. Christensen, Y. Ding, L. Frandsen, K. Rottwitt, and Leif Oxenløwe, "Unidirectional frequency conversion in microring resonators for on-chip frequency-multiplexed single-photon sources," New J. Phys. \textbf{21}, 033037 (2019).

\bibitem{Lacava}
C. Lacava, M. A. Ettabib, T. D. Bucio, G. Sharp, A. Z. Khokhar, Y. Jung, M. Sorel, F. Gardes, D. J. Richardson, P. Petropoulos, and F.
Parmigiani, "Intermodal Bragg-Scattering Four Wave Mixing in Silicon Waveguides," J. Lightwave Techol. \textbf{37}, 1680 (2019).

\bibitem{Sven}
S. Ramelow, A. Farsi, Z. Vernon, S. Clemmen, X. Ji, J. E. Sipe, M. Liscidini, M. Lipson, and A. L. Gaeta, "Strong Nonlinear Coupling in a $\mathrm{Si}_{3}\mathrm{N}_{4}$ Ring Resonator,"
Phys. Rev. Lett. \textbf{122}, 153906 (2019).

\bibitem{Hong Tang}
X. Guo, C.-L. Zou, H. Jung, and H. Tang, "On-Chip Strong Coupling and Efficient Frequency Conversion between Telecom and Visible Optical Modes," 
Phys. Rev. Lett. \textbf{117}, 123902 (2016).


\bibitem{Adibi}
E. S. Hosseini, S. Yegnanarayanan, A. H. Atabaki, M. Soltani, and A. Adibi, "Systematic design and fabrication of high-Q single-mode pulley-coupled planar silicon nitride microdisk resonators at visible wavelengths," Opt. Express \textbf{18}, 2127-2136 (2010).

\bibitem{thermal_annealing}
M. Blasco, S. Dacunha, C. Dominguez, and J. Faneca, "Silicon nitride stoichiometry tuning for visible photonic integrated components," Appl. Phys. Lett. \textbf{124}, 221104 (2024).

\bibitem{nitrogen_rich}
J. Smith, J. Monroy-Ruz, J. Rarity, K. Balram, "Single photon emission and single spin coherence of a nitrogen vacancy center encapsulated in silicon nitride," Appl. Phys. Lett. \textbf{116}, 134001 (2020).

\bibitem{silicon_nanoparticles}
N. Daldosso and L. Pavesi, Laser Photonics Rev. 3, 508 (2009).

\bibitem{band_tail}
J. Kistner, X. Chen, Y. Weng, H. P. Strunk, M. B. Schubert, J. H. Werner, "Photoluminescence from Silicon Nitride - No Quantum Effect," J. Appl. Phys. \textbf{110}, 023520 (2011).

\bibitem{Ryota}
R. Katsumi, K. Takada, S. Naruse, K. Kawai, D. Sato, T. Hizawa, T. Yatsui, "Hybrid integration of ensemble nitrogen-vacancy centers in single-crystal diamond based on pick-flip-and-place transfer printing," Appl. Phys. Lett. \textbf{123}, 111108 (2023).

\end{thebibliography}

\begin{thebibliography}{1}
\newcommand{\enquote}[1]{``#1''}

\bibitem{Yun Thesis}
Y. Zhao, "Nonlinear Photonics for Room-Temperature Quantum Metrology and Information Processing," PhD Thesis, Columbia University (2022).

\bibitem{Sipe}
Z. Vernon, M. Liscidini, and J.E. Sipe, "Quantum frequency conversion and strong coupling of photonic modes using
four-wave mixing in integrated microresonators," Phys. Rev. A \textbf{94}, 023810 (2016).

\bibitem{Yun Theory}
Y. Zhao, J. Jang, Y. Okawachi, and A. L. Gaeta, "Theory of $\chi$(2)-microresonator-based frequency conversion," Opt. Lett. \textbf{46}, 5393-5396 (2021).

\end{thebibliography}
\end{document}